# IoT Smart Plant Monitoring, Watering and Security System

U.H.D. Thinura Nethpiya Ariyaratne
Undergraduate, Faculty of Computing,
Sri Lanka Institute of Information Technology,
Malabe, Sri Lanka
it19198200@my.sliit.lk

V. Diyon Yasaswin Vitharana
Undergraduate, Faculty of Computing,
Sri Lanka Institute of Information Technology,
Malabe, Sri Lanka
it19196770@my.sliit.lk

L.H. Don Ranul Deelaka
Undergraduate, Faculty of Computing,
Sri Lanka Institute of Information Technology,
Malabe, Sri Lanka
it19198446@my.sliit.lk

H.M. Sumudu Maduranga Herath
Undergraduate, Faculty of Computing,
Sri Lanka Institute of Information Technology,
Malabe, Sri Lanka
it19199368@my.sliit.lk

## 1 Abstract

Interest in home gardening has burgeoned since governments around the world-imposed lockdowns to suppress the spread of COVID-19. Nowadays, most families start to do gardening during this lockdown season because they can grow vegetables and fruits or any other plants that they want in their day-to-day life. So, they can survive without spending money on online grocery shopping for fruits and vegetables during this lockdown season. In Sri Lanka, home gardening was a trend during the past couple of months due to this pandemic. Most of the families were trying to do gardening for their needs. But the problem is, nowadays the government is trying to release those restrictions to start day-to-day work in Sri Lanka. With this situation, people are starting to do their jobs and they do not have time to spend in their gardens continuing their gardening. We thought about this problem and tried to find a solution to continue the gardening work while doing their jobs. The major concern is people cannot monitor their plants every time and protect their garden. So, we decided to automate the garden work. With our new solution, gardeners can monitor some important factors like the plant's healthiness, soil moisture level, air humidity level, and the surrounding temperature and water their garden from anywhere in the world at any time by using our app. Plant health has a significant impact on plant development, production, and quality of agricultural goods. The goal of this study is to create an automated system that can identify the presence of illness in plants based on variations in plant leaf health state is created utilizing sensors such as temperature, humidity, and color. Users can also make a watering schedule so that our unit will automatically water the plants by considering several factors and they do not have to manually water the plants every day. As an added feature, we also developed a security mechanism that will detect the movements in the surrounding of plants and scare away the animals while alerting the user if required.

## 2 Introduction

Since governments throughout the world implemented lockdowns to combat the spread of COVID-19, there has been a surge in interest in home gardening. Most families nowadays begin gardening during this season since they may grow whatever veggies, fruits, or other plants they desire in their daily lives. As a result, they will be able to make it through this period without having to spend money on online grocery shopping for fruits and vegetables. Due to the epidemic in Sri Lanka, home gardening has been popular in recent months. Most of the households were attempting to grow for their purposes.

However, the government is currently attempting to lift such limitations to allow people to return to regular work in Sri Lanka. People are starting to do their jobs because of this scenario, and they do not have time to continue planting in their gardens. We discussed the issue and attempted to develop a solution that would allow them to continue working while doing so. The main issue is that individuals cannot constantly watch their plants to safeguard their gardens. As a result, we decided to automate the garden labor.

Gardeners may use our innovative solution to monitor critical parameters such as plant health and illness, soil moisture level, air humidity level, and surrounding temperature, as well as water their garden from anywhere in the globe at any time.

The current approach is based on observation with the naked eye, which is a time-consuming process. To detect plant disease at an early stage, automatic detection of plant disease can be used. Farmers have utilized a variety of disease control techniques daily to prevent plant illnesses.

Plant disease has a substantial influence on plant growth, productivity, and agricultural product quality. The objective of this research is to develop an automated system that can detect the health of plants. An automated health detection system is built based on fluctuations in plant leaf health conditions using sensors such as temperature, humidity, and color.

The illness starts on the plant leaves in most cases. As a result, we have considered the detection of plant health on leaves in the proposed study. Temperature, humidity, and color variations can be used to distinguish between normal and damaged plant leaves.

Users can manually water the plants by checking the soil moisture level at any time and from anywhere in the world, the user may also create a watering schedule so that our device will automatically water the plants based on several criteria, eliminating the need to water the plants manually every day.

We also create a security mechanism that will detect movements in the environment of the plants and inform the user if they are discovered. We'll create an IoT gadget that can be installed in a garden and operated by an app or a web platform.

## 3    Literature Review

### 3.1    Smartphone irrigation sensor

To be used in crops, an automatic irrigation sensor was devised and built. [1] The sensor captures and processes digital pictures of the soil around the crop's root zone with a smartphone and calculates the water content visually. The sensor is housed in a room with regulated lighting and buried at the plant's root level. The smartphone's processing and connection components, such as the digital camera and the Wi-Fi network, were controlled directly by an Android App. The smartphone is activated by the mobile App, which wakes it up according to user-defined settings. Through an anti-reflective glass window, the built-in camera takes a picture of the soil., and an RGB to a gray procedure is used to estimate the ratio between the image's moist and dry areas. After established a

Wi-Fi connection, the ratio is sent to a gateway for control of an irrigation water pump through a router node.

## 3.2 Predicting the extent of wildfires using remotely sensed soil moisture and temperature trends

Weather patterns changes reveal an increase in temperatures, as well as a growth in the duration and frequency of drought, resulting in more severe wildfires, which endangers both the environment and human life. [2] Improved wildfire prediction technologies are critical in this setting. This research explored the significance of remotely sensed soil moisture data as a crucial variable in the climate-wildfires association. This research focuses on fires that occurred in the Iberian Peninsula between 2010 and 2014. When investigating their prior-to-occurrence surface moisture temperature conditions, researcher utilized SMOS-derived soil moisture data and ERA-Interim land surface temperature reanalysis.

## 3.3 Wireless sensor network based automated irrigation and crop field monitoring system

Sensors that can be used wirelessly for agricultural purposes. In order to reduce water usage, a network-based automated watering system is utilized. [3] In the agricultural field, A wireless sensor network comprising soil moisture and temperature sensors is used in the system. To handle sensor data, the Zigbee protocol was utilized, and an algorithm using sensor threshold values was used to regulate the water amount programmed on a microcontroller for irrigation. For data examination, A solar panel powers the gadget, which also has a cellular internet interface.

Using image processing techniques, to monitor the disease area, a wireless camera is installed in the agricultural field. The technology is low-cost and energy-independent, making it ideal for water-scarce and geographically isolated locations.

## 3.4 "Project Haritha" - an automated irrigation system for home gardens

In an urban setting, even the upkeep of a tiny garden may become tiresome. The need of the hour is for a completely automated system that maximizes the utilization of energy and water resources. As a consequence, a multi-mode, extremely energy-efficient control for an automated irrigation system has been developed. was designed and implemented. To irrigate a targeted region, the system employs an in-situ soil moisture [4]potential sensor and programmed data. The soil moisture content is monitored using a microcontroller-based data collection and dissemination system. During a system failure, an embedded GSM module gives vital information to the user. The recommended microcontroller-based system was written and tested to see how well it performed.

## 3.5 Automated irrigation system using a wireless sensor network and GPRS module

To improve water utilization for crops, an automated irrigation system was created. [5] The system consists of a distributed wireless network of soil moisture and temperature sensors in the root zone of plants. A gateway device also manages sensor data, triggers actuators, and delivers information to a web application. To manage water amount, an algorithm containing temperature and soil moisture threshold values was created and implemented into a gateway based on a microcontroller Panels for photovoltaics powered the system, A simultaneous connection based on a cellular Internet interface allowed data inspection and irrigation planning to be configured via a web page.

### 3.6 The Design and Research on Intelligent Fertigation System

Crop fertilization is uneven, resulting in either too much or too little fertilization, and the concentration cannot be effectively regulated, which is a serious but typical problem. [6] Accordingly, an intelligent fertilization device has been developed that can automatically water and fertilize, as well as inject and combine fertilizer. Three control algorithms are presented in the paper: fertilizer application control algorithm; fertilizer injection and mixing; and a system priority algorithm. It also describes the system structure and the design of the piping system. This system has a good quality for EC and pH adjustment, a steady performance, and is highly practical, according to the testing findings.

## 4 Methodology

This project is designed to monitor, water, and provide security for the plants in the garden. For this project, we are selecting the Turmeric plant as our testing garden plant. In this situation, a lot of people are interested in planting Turmeric. Also, this is an IoT-based project. It provides you to monitor and control the system from anywhere in the world. This system consists of different sensors to get data from the plants. They are soil moisture sensor, temperature and humidity sensor, color sensor, and motion sensor.

These sensors are used to take the data from the plants by monitoring it and make some decisions according to the inputs. This system has four main operations. They are as follow,

**Operation 1**
- Obtaining the soil moisture level of surrounding to water the plants according to the moisture level, the user defines time slot, or manually user input

**Operation 2**
- Obtaining the color values of the leaves, temperature, and humidity of surrounding of the plant to monitor the health.

**Operation 3**
- Detecting the motion around the plants.

**Operation 4**
- Obtaining the temperature and humidity levels of the surrounding and displaying it to the user

We use a pre-defined dataset to check the important factors of the plant. The table 2.1 represents that dataset.

| Measuring factor | | Low level | Mid-level | High Level |
|---|---|---|---|---|
| Color of the leaves | Red | {5,645,820,1110} | - | {9,698,1095,1350} |
| | Green | {4,770,1050,1550} | - | {6,835,1565,2245} |
| | Blue | {4,1090,1698,2490} | - | {5,1207,1290,2793} |
| Humidity of plant | | 60% | - | 80% |
| Temperature of plant | | 20 °C | - | 35 °C |
| Temperature of the surrounding | | - | - | 40 °C |
| Humidity of the surrounding | | 30% | - | - |
| Moisture level | | 40% | 70% | 100% |

*Table 4.1 Description about measuring factors and their levels*

When discussing obtaining soil moisture level, temperature level, and humidity level of surroundings, a soil moisture sensor is used to extract those factors from the soil. [7] [8]Not only that we have divided the range of moisture into several levels which are minimum level, average level, and high level as shown in Table 1. When choosing those values, we are using loam soil to grow our test plant.

Users can manually water the plants using the mobile app and the system can water the plants automatically at the user define time slots when the soil moisture is under average level. When moisture levels come to a high level, the system will automatically alert the user to "Do not water the plant until the water level comes between a minimum and average level". This whole process is briefly shown in Figure 2.1. When the

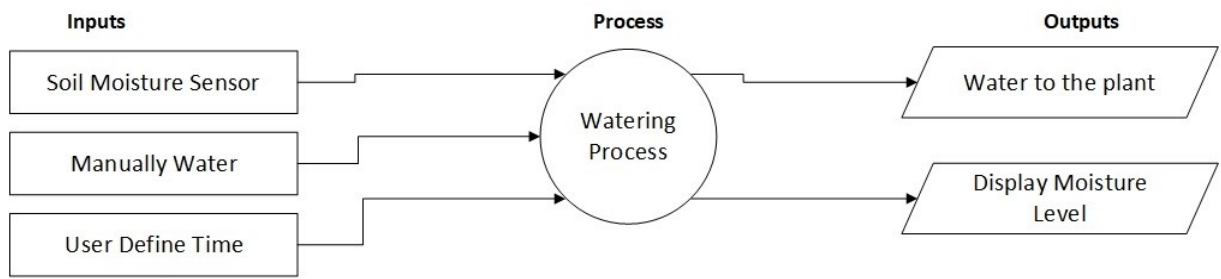

*Figure 4.1 : Describing the plant watering process*

Because most of the home gardening plants used loam soil. When the water level is under the minimum level, the system will automatically water the plants and maintain the moisture up to the minimum level.

surrounding temperature goes high, the system will automatically alert the temperature is high. Also, the humidity will go down, system will automatically alert the humidity is low.

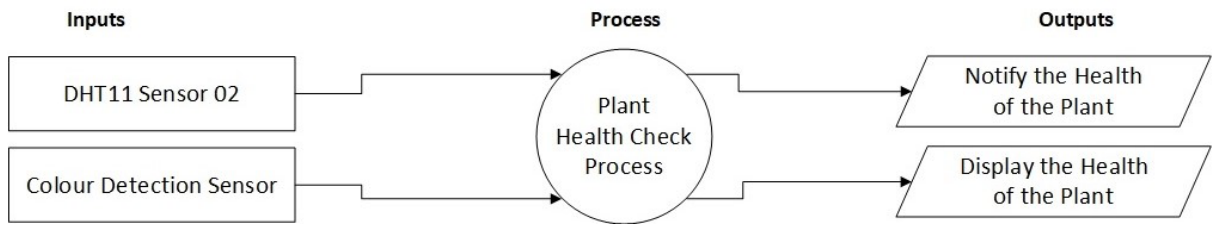

*Figure 4.2 : Describing the plant health check process*

When discussing plant health identification, there are two sensors used to collect the data to make that decision which would be the temperature and humidity sensor and the color sensor as mentioned in Figure 2.2. So, the color sensor is used to check the color difference of the leaves. [9] These 3 facts are required to identify the health of the plant. Three factors are required to evaluate the plant's health. The first factor is the temperature, we obtain the temperature of the plant using the temperature and humidity sensor then we process is with our algorithm. [10] During that we check the value according to the predefined range of temperature data as shown below,

```
if (min temperature < temperature< max temperature)
    Pass healthy
else
    Pass not healthy
```

The second factor is humidity. We obtain the humidity level of the plant using the temperature and humidity sensor. [11] Then the value is checked according to the predefined range of humidity data and shows the health of the plant to the user. There is an algorithm that we are using as shown below,

```
if (min humidity < humidity< max humidity)
    Pass healthy
else
    Pass not healthy
```

The third factor is, identifying the plant's health using the color difference of the leaves. Using color sensor, we obtain the RGB values of the color of the leaf and compare with the predefined range of values in the dataset to determine whether the leaf is healthy or diseased and show to the user. We used multiple color values as shown in Table 1 because in the same plant there are baby and normal leaves can excite. Then we need to consider all types of leaves. When collecting the unhealthy color values, we consider multiple

```
if (min RED value < RED value< max RED value)
    Pass not ok
else
    Pass ok

if (min BLUE value < BLUE value< max BLUE value)
    Pass not ok
else
    Pass ok

if (min GREEN value < GREEN value< max GREEN value)
    Pass not ok
else
    Pass ok

if (RED value ok and BLUE value ok and GREEN value ok)
    Pass healthy
else
    Pass not healthy
```

unhealthy leaves and place the color sensor in different positions and angles.

Finally, we separately check these factors and show the plant healthy as if one factor is ok, health level is shown as 30% for two factors are ok, health level is shown as 60% and if all three factors are ok, healthy level is shown as 90%. A certain plant is considered as healthy if at least two of the above three conditions are satisfied.

When starting a plantation at home, another major issue is animal attacks. Birds, squirrels, rats, home pets like animals can come to the plantation area and destroy the plants. This system has a mechanism to avoid attacks like this. It has a motion detection sensor to detect motions around the plants and when the system detects any motion around the area, it turns on the security buzzer and lights as mentioned in Figure 2.3. This security system can be manually on/off by using the mobile app.

when the temperature is increased, or humidity is decreased. We predetermined the high temperature and low humidity of the surrounding as shown in Table 1.

```
if (surround temperature >= high level)
{
    Display surrounding temperature is high
}

if (surround humidity <= high humidity)
{
    Display surrounding humidity is low
}
```

Blynk platform is a cloud platform that is used to connect our IoT system with the mobile app. It has an API to display the parameters that is collected from the system. Users can know the moisture level of the soil, temperature, and humidity of the surroundings,

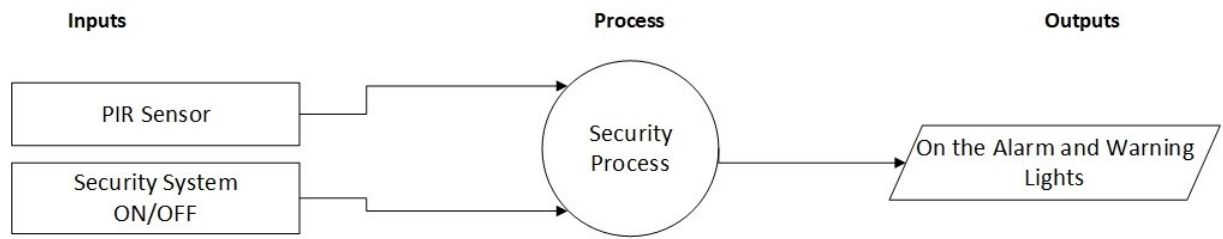

Figure 4.3 : Describing the plant security process

We use another temperature and humidity sensor to collect the temperature and humidity of the surrounding and show it to the user to get a brief idea about the surrounding conditions of the plant. Also, according to the collected data we indicate the user and the health level of the plant.

Also, it has dedicated buttons for manually watering, on/off the security system, and enter the time slot to water the plants automatically daily.

# 5 Testing and Evaluation

## 5.1 Test Cases

### 5.1.1 Test Case 1 - Testing the soil moisture level identification [12]

a) Placing a turmeric plant in a pot with loam soil.
b) Connecting the sensor to the board and initializing it.
c) Keeping it for a while without adding water in order to get a dry soil. (Low level)
d) Placing the soil moisture sensor and obtaining the value from it.
e) Preparing a cup of water and placing the soil moisture sensor in it.
f) Obtaining sensor reading and noting down the value. (High level)

With these data we can determine the soil moisture ranges; low, mid and high levels.

### 5.1.2 Test Case 2 - Testing the water flow mechanism

**A. Testing the relay module** [13]
a) Connecting the relay module to the board and initializing it.
b) Connecting a 12V bulb with the relay module.
c) Testing the relay module by turning the bulb on and off using it.

**B. Testing the Solenoid Valve** [14]
a) Preparing a water collection container.
b) Connecting the Solenoid valve with the relay module to the board.
c) Testing the opening and closing events of the valve using the relay module.
d) Connecting the water input and the water output pipes to and from the valve.
e) Testing the water flow.

**C. Testing the integrated water flow and the soil moisture identification mechanisms**
a) Connecting all the three items to the board and initializing them.
b) Placing a dry soil amount in the turmeric pot to check whether it is displayed in the app and the water flow will be high as the soil moisture level will be less.
c) Placing a wet soil amount in the turmeric pot to check whether it is displayed in the app and the water flow will not happen as the soil moisture level is high.

With these data, we can check whether the water flow mechanism is working correctly.

### 5.1.3 Test Case 3 – Testing the plant health identification mechanism

**A. Testing the humidity sensor** [15]
a) Connecting the humidity sensor to the board and initializing it.
b) Placing it in a sunny and dry location. Obtain the readings. (Low Humidity)
c) Placing it in indoors and sprinkling some water to the air. Obtaining the readings. (High Humidity)
d) Determining the high and low values of humidity.

**B. Testing the temperature sensor** [15]
a) Connecting the temperature sensor to the board and initializing it.
b) Placing it in a sunny location for some time and obtaining the readings. (High level)
c) Placing it in a cool area and obtaining the readings. (Low level)
d) Determining the temperature level ranges.

C. **Testing the color sensor** [16]
a) Connecting the color sensor to the board and initializing it.
b) Obtaining red, green, blue colored paper pieces.
c) Pointing the sensor for those three pieces and checking whether the colors are identified correctly.
d) Determining the values and the color ranges.

D. **Integrating the humidity, temperature, and color sensors**
a) Connecting all the three items to the board.
b) Checking whether the user is notified when there is a change in humidity is identified in the soil by changing the plant health status.
c) Checking whether the user is notified when there is a change in temperature is identified in the soil by changing the plant health status.
d) Checking whether the user is notified when there is a change of color is identified in the plant leaf by changing the plant health status.

### 5.1.4 Test Case 4 – Testing the plant security mechanism

a) Connecting the motion detection sensor to the board and initializing it.
b) Connecting the buzzer and LED lights to the board and initializing them.
c) Obtaining the output values from the motion sensor when a motion is detected and when there is no motion detected.

Configuring the buzzer and the LED bulbs to turn on and off when motion is detected or not.

### 5.2 Evaluation

- By comparing with the actual readings and the automated readings that we obtained.
- By comparing with the outcomes of the other similar research done.

## 6 Conclusion

As a result, the "IoT Smart Plant Monitoring, Watering and Security System" has been successfully planned and constructed. It was created by combining the features of all of the hardware components used. Every module's presence has been carefully considered and positioned, resulting in the best possible operation of the unit. The system was thoroughly tested to ensure that it will run on its own. The moisture sensors monitor the water content (moisture level) of the soil of turmeric plant. The moisture sensor provides a signal to the microcontroller when the moisture level falls below the specified level, causing the water pump to come on and feed water to the appropriate plant. The system automatically comes to a halt when the required moisture level is attained, and then the water pump is switched off. Other than that, if the humidity and temperature levels are changed, it is alert the user through the app. Users have the ability to monitor and control the units from anywhere in the world at any time. Also, they have the opportunity to make a watering schedule according to their preference so that can reduce the manual human interactions and water the plant automatically by checking the moisture levels of the soil. Users can also manually water the plant simply by using the mobile app which controls the IOT device. As the special function of the IoT device, the device can check the plant's healthiness by checking the color of plant's leaves, plant's surrounding temperature and the humidity. Also, we used several pre-determined datasets to check whether the plant is healthy or not with the help of the above factors. As an added feature, we implemented a security mechanism that is detect motions of animals by monitoring the surrounding of the plant and alert the users if required. If an unusual behavior has been detected, the buzzer and the lights are started to scare away the animal. Accordingly, the entire system's operation has been carefully tested and is deemed to work well.

# 7  Future work

According to our project, we implemented our IOT device for a one plant as a prototype. As the further work, we can develop our IoT device for home garden which can check all the plants and the soil. Also, we can develop this IOT device which can use for green houses. Then the green house will be automated, and it will help to get the expect outcome easily or monitor the green house from anywhere in the world. As the industrial level, we can develop our IOT device which can use for farmers to monitor their farm form their mobile phones.